\documentclass[aps,prc,superscriptaddress,twoside,twocolumn,nofootinbib,showpacs,floatfix]{revtex4-1}
\usepackage{amsmath,amssymb}
\usepackage{graphicx}
\usepackage{subfigure}
\usepackage{ulem}
\usepackage[colorlinks]{hyperref}
\usepackage[usenames,dvipsnames]{color}
\pdfpagewidth=\paperwidth
\pdfpageheight=\paperheight
\allowdisplaybreaks
\usepackage{newtxtext, newtxmath}
\usepackage{braket,bm}
\usepackage{multirow}
\usepackage{enumitem}
\usepackage[vcentermath]{youngtab}
\usepackage{appendix}
\usepackage{gensymb}

\begin{document}

\title{\boldmath Hidden charm pentaquark states in a diquark model}

\author{Pan-Pan Shi}
\affiliation{School of Nuclear Science and Technology, University of Chinese Academy of Sciences, Beijing 101408, China}
\affiliation{CAS Key Laboratory of Theoretical Physics, Institute of Theoretical Physics, Chinese Academy of Sciences, Beijing 100190, China}

\author{Fei Huang}
\affiliation{School of Nuclear Science and Technology, University of Chinese Academy of Sciences, Beijing 101408, China}

\author{Wen-Ling Wang}
\email[Email: ]{wangwenling@buaa.edu.cn}
\affiliation{School of Physics, Beihang University, Beijing 100191, China}

\date{\today}

\begin{abstract}
The mass spectrum of hidden charm pentaquark states composed of two diquarks and an antiquark are calculated by use of an effective Hamiltonian which includes explicitly the spin, color, and flavor dependent interactions. The results show that the $P_c(4312)^+$ and $P_c(4440)^+$ states could be explained as hidden charm pentaquark states with isospin and spin-parity $IJ^P=1/2\left(3/2^-\right)$, the $P_c(4457)^+$ state could be explained as a hidden charm pentaquark state with $IJ^P=1/2\left(5/2^-\right)$, and the $P_{cs}(4459)^+$ state could be explained as a hidden charm pentaquark state with $IJ^P=0\left(1/2^-\right)$ or $0\left(3/2^-\right)$. Predications for the masses of other possible pentaquark states are also given, and the possible decay channels of these hidden charm pentaquark states are discussed.
\end{abstract}



\maketitle

\section{Introduction}

In 2015, two pentaquark candidates, $P_c(4380)^+$ and $P_c(4450)^+$ have been reported by the LHCb Collaboration in the $\Lambda_b \rightarrow KJ/\psi p$ process \cite{Aaij:2015tga,Aaij:2016phn,Aaij:2016ymb}. The $P_c(4380)^+$ and $P_c(4450)^+$ have masses $M=4380\pm 8 \pm 29$ MeV and $M=4449.8 \pm 1.7 \pm 2.5$ MeV, decay widths $\Gamma= 205 \pm 18\pm 86$ MeV and $\Gamma= 39 \pm 5 \pm 19$ MeV, and spins $J=3/2$ and $J=5/2$, respectively. Moreover, these two states have opposite parity. In 2019, the $P_c(4312)^+$ with a mass $M=4311.9 \pm 0.7^{+6.8}_{-0.6}$ MeV and a width $\Gamma=9.8 \pm 2.7^{+3.7}_{-4.5}$ MeV has been observed by the LHCb Collaboration, and the formerly observed $P_c(4450)^+$ has been reported to consist of two overlapping peaks, the $P_c(4440)^+$ with a mass $M=4440.3 \pm 1.3^{+4.1}_{-4.7}$ MeV and a width $\Gamma=20.6 \pm 4.9^{+8.7}_{-10.1}$ MeV, and the $P_c(4457)^+$ with a mass $M=4457.3 \pm 0.6^{+4.1}_{-1.7}$ MeV and a width $\Gamma=6.4 \pm 2.0^{+5.7}_{-1.9}$ MeV \cite{Aaij:2019vzc}. Very recently, the LHCb Collaboration reported the $P_{cs}(4459)$ state with a mass $M=4458.8 \pm 2.9^{+4.7}_{-1.1}$ MeV and a width $\Gamma=17.3 \pm 6.5^{+8.0}_{-5.7}$ MeV in the reaction $\Xi_b \to J/\psi K^- \Lambda$  \cite{Aaij:2020}.

Theoretically, the $P_c$ states have been attracting a lot of interests in the hadron physics community because of their exotic structures. The $P_c$ states were observed in the $J/\psi p$ channel, so their quark constituents possibly are $uudc\bar{c}$. In order to investigate the nature of those $P_c$ states, many theoretical models were put forward. The most commonly used models are baryon-meson molecule model \cite{Wu:2010jy,Yang2012,Chen2016a,Shimizu2016,Yamaguchi:2016ote,Eides2016,Azizi2017,Guo2019,Liu:2019tjn,Xiao2019,Du2020,Guo:2019kdc,Fernandez-Ramirez:2019koa,Du:2021fmf,Dong:2021juy}, kinematical effect \cite{Mikhasenko2015,Guo2015,Liu2016,Bayar:2016ftu}, and constituent quark model \cite{Park2017,Santopinto2017,Deng2017,Zhu2019,Dong2020}. The studies of the $P_c$ states were reviewed in Refs.~\cite{Esposito2017,Lebed2017,Guo2018,Liu2019}. In addition, the diquark model was also used to discuss the $P_c$ states, e.g. the mass spectra of the $P_c$ states were calculated within the triquark-diquark model \cite{Zhu2016a,Lebed2017a,Giron:2019bcs,Ali:2019clg} and diquark-diquark-antiquark model \cite{Maiani2015,Anisovich2015,Lebed2015,Li2015,Wang2016,Chen2016,Ali2016,Ghosh2017,Ali:2019npk,Semenova2019}, respectively. 

Even the $P_c$ states have been widely investigated in diquark models, no conclusive answers have been obtained so far for the quantum numbers of these states. For example, the $P_c(4380)^+$ was reported to have spin-parity $J^P=3/2^-$ in Refs.~\cite{Maiani2015,Lebed2015,Ali2016,Zhu2016a}, and $J^P=3/2^+$ in Ref.~\cite{Anisovich2015}; the $P_c(4312)^+$, $P_c(4440)^+$, and $P_c(4457)^+$ were reported to have spin-parity $J^P=3/2^-$, $3/2^+$, and $5/2^+$, respectively, in Ref.~\cite{Ali:2019npk,Ali:2019clg}, and have $J^P=1/2^-$, $1/2^-$, and $3/2^-$, respectively, in Ref.~\cite{Semenova2019}. 

In this work, we study the spectrum of hidden charm pentaquark states in a diquark model. Concretely, we construct the wave functions for all possible $qqqc\bar{c}$, $qqsc\bar{c}$, $qssc\bar{c}$, $sssc\bar{c}$ configurations ($q=u,d$) in a diquark-diquark-antiquark model, and calculate the masses of these states by use of an effective Lagrangian, which includes explicitly the color, spin, and flavor dependent interactions. We discuss the assignments of the quantum numbers for the $P_c$ and $P_{cs}$ states, and give predictions for the masses of the flavor partners of the $P_c$ and $P_{cs}$ states and of other possible pentaquark states.

The paper is organized as follows. In Sec.~\ref{Sect:wave function}, the wave functions for the hidden charm pentaquark states are introduced. In Sec.~\ref{Sect:Mass_formula}, the effective Hamiltonian and the mass formulas for the pentaquark states are given. In Sec.~\ref{Sect:Numerical_result}, the masses of the hidden charm pentaquark states are calculated, and some discussions are presented. In Sec.~\ref{Sec:decay}, the possible decay channels of hidden charm pentaquark states are discussed. Finally, a summary is given In Sec.~\ref{Sect:Summary}.

\section{Wave functions of the hidden charm pentaquark states}  \label{Sect:wave function}

In color space, it is well known that a diquark is in the $\bar{{\bf{3}}}_c$ representation of the SU$_c$(3) group \cite{Shi2019}. Considering that the whole multiquark system should be in a color singlet, the color wave functions for the diquark-diquark-antiquark can be constructed as
\begin{align}
\left\{ \left\{~\yng(1,1)~\otimes~\yng(1,1)~\right\}_{\scriptsize~\yng(2,1,1)}~\otimes~\yng(1,1)~\right\}_{\scriptsize~\yng(2,2,2)}.         \label{Eq:color_wave_function}
\end{align}

If we consider only the states with orbital angular momentum $L=0$, then in spin-flavor space, two quarks in a diquark should be symmetric, since they are already antisymmetric in color space. The possible diquark configurations are (0, $\bar{{\bf{3}}}_f$, $\bar{{\bf{3}}}_c$) and (1, ${\bf{6}}_f,$ $\bar{{\bf{3}}}_c$) \cite{Shi2019}, where in each parenthesis, the first number depicts the spin of the diquark, the second and third numbers represent the irreducible representations of the two-quark system in flavor SU$_f$(3) and color SU$_c$(3) spaces, respectively.

As mentioned above, in flavor space, a diquark could be in $\bar{{\bf{3}}}_f$ or ${\bf{6}}_f$ representation of the SU$_f$(3) group. For the hidden charm pentaquark state composed of two diquarks and an antiquark, the flavor wave function for the three light quarks can be constructed as
\begin{align}
& \left\{\left\{~\yng(1)~\otimes~\yng(1)~\right\}_{\scriptsize~\yng(1,1)}~\otimes~\yng(1)~\right\}_{\scriptsize~\yng(1,1,1)},   \\
& \left\{\left\{~\yng(1)~\otimes~\yng(1)~\right\}_{\scriptsize~\yng(1,1)}~\otimes~\yng(1)~\right\}_{\scriptsize~\yng(2,1)},  \\[3pt]
& \left\{\left\{~\yng(1)~\otimes~\yng(1)~\right\}_{\scriptsize~\yng(2)}~\otimes~\yng(1)~\right\}_{\scriptsize~\yng(3)},   \\[4pt]
& \left\{\left\{~\yng(1)~\otimes~\yng(1)~\right\}_{\scriptsize~\yng(2)}~\otimes~\yng(1)~\right\}_{\scriptsize~\yng(2,1)}.
\label{Eq:flavor_function}
\end{align}
In the following parts of the paper, we concisely denote these four states as ${\bf{1}}_f$, ${\bf{8}}^{\bf{A}}_f$, ${\bf{10}}_f$, and ${\bf{8}}^{\bf{S}}_f$, respectively.

When the three light quarks are in flavor ${\bf{1}}_f$, ${\bf{8}}^{\bf{A}}_f$, ${\bf{10}}_f$, and ${\bf{8}}^{\bf{S}}_f$ configurations, the explicit flavor wave functions of the whole pentaquark system read
\begin{align}
& \Psi({\bf{1}}_f,0,-1) = \sqrt\frac{1}{3} [qq] (sc) \bar{c} + \sqrt\frac{2}{3} \left( [qs] (qc) \right)_0\bar{c},
\label{Eq:flavor_singlet}    \\[3pt]
& \Psi({\bf{8}}^{\bf{A}}_f,1/2,0) = [qq] (qc) \bar{c},  \label{Eq:octet_q_A}  \\[3pt]
& \Psi({\bf{8}}^{\bf{A}}_f,1/2,-2) = [qs] (sc) \bar{c},   \label{Eq:octet_2s_A}  \\[3pt]
& \Psi({\bf{8}}^{\bf{A}}_f,0,-1) = -\sqrt{\frac{2}{3}} [qq] (sc) \bar{c} - \sqrt\frac{1}{3} \left( [qs] (qc) \right)_0\bar{c},    \label{Eq:octet_s_0_A} \\[3pt]
& \Psi({\bf{8}}^{\bf{A}}_f,1,-1) = \left( [qs] (qc) \right)_0 \bar{c},  \label{Eq:octet_s_1_A} \\[3pt]
& \Psi({\bf{10}}_f,3/2,0) = \left( \{qq\} (qc) \right)_{3/2} \bar{c},  \label{Eq:decuplet_q} \\[3pt]
& \Psi({\bf{10}}_f,1,-1) = \sqrt\frac{2}{3}\left( \{qs\} (qc) \right)_1\bar{c} + \sqrt\frac{1}{3} \{qq\}(sc)\bar{c}, \label{Eq:decuplet_s}  \\[3pt]
& \Psi({\bf{10}}_f,1/2,-2) = \sqrt\frac{2}{3} \{qs\} (sc) \bar{c} + \sqrt\frac{1}{3} \{ss\} (qc) \bar{c},\label{Eq:decuplet_2s}   \\[3pt]
& \Psi({\bf{10}}_f,0,-3) = \{ss\} (sc) \bar{c}, \label{Eq:decuplet_3s}  \\[3pt]
&\Psi({\bf{8}}^{\bf{S}}_f,1/2,0) = -\left(\left\{qq\right\} (qc) \right)_{1/2} \bar{c},   \label{Eq:octet_q_S}  \\[3pt]
&\Psi({\bf{8}}^{\bf{S}}_f,0,-1) = -\left(\left\{qs\right\} (qc) \right)_0 \bar{c},  \label{Eq:octet_s_0_S} \\[3pt]
&\Psi({\bf{8}}^{\bf{S}}_f,1,-1) = \sqrt{\frac{2}{3}} \left\{qq\right\} (sc) \bar{c} -\sqrt\frac{1}{3} \left(\left\{qs\right\} (qc) \right)_1 \bar{c},  \label{Eq:octet_s_1_S}  \\[3pt]
&\Psi({\bf{8}}^{\bf{S}}_f,1/2,-2) = \sqrt\frac{1}{3} \{qs\} (sc) \bar{c} - \sqrt{\frac{2}{3}} \{ss\}(qc) \bar{c}. \label{Eq:octet_2s_S}
\end{align}
Here the second and third indices in parenthesis on the left hand side denote, respectively, the isospin and strangeness of the whole system. The symbol $q$ represents the $u, d$ quark. The brackets $``\left\{~\right\}"$ and $``\left[~\right]"$ for the first diquark denote that the two quarks inside this diquark are symmetric and antisymmetric, respectively, in SU(3) flavor space. The spin for this diquark is, then, $1$ and $0$, respectively, for these two cases to ensure that the two quarks are symmetric in spin-flavor space. For the second diquark, the bracket $``(~)"$ denotes that the flavor symmetry of the two quarks inside this diquark is not definite. Actually, the flavor symmetry of this diquark will not be clear until the spin of this diquark is given, as the only requirement of the symmetry is that the two quarks are symmetric in spin-flavor space. On the right hand side of the above equations, we also give explicitly the total isospin of the two diquarks in case that there are more than one possibility for the total isospin when couple the two diquarks.

In spin space, the wave function of the pentaquark system can be denoted as
\begin{align}
\Ket{S_{12},S_{34},S_{1234},S} \equiv \left\{\left[\left(s_1s_2\right)_{S_{12}} \left(s_3s_4\right)_{S_{34}} \right]_{S_{1234}} s_{\bar 5}\right\}_S,    \label{eq:wf_spin}
\end{align}
where $S_{12}=1$ and $0$, respectively, for symmetric and antisymmetric flavor wave functions for the first diquark. The spin for the second diquark $S_{34}=1$ or $0$, and consequently, the flavor symmetry for this diquark should be symmetric or antisymmetric, respectively.

\section{Mass formulas for hidden charm pentaquark states} \label{Sect:Mass_formula}

\subsection{The effective Hamiltonian} \label{subsect:Hamiltonian}

\begin{table*}[htb]
\caption{\label{Tab:Spin_matrix} Spin matrix elements for pentaquark states. The spin wave function $\Ket{S_{12},S_{34},S_{1234},S}$ is defined in Eq.~(\ref{eq:wf_spin}).}
\renewcommand{\arraystretch}{1.3}
\begin{tabular*}{\textwidth}{@{\extracolsep\fill}ccccccccccc}
\hline\hline                                           
  &  $S_1\cdot S_2$  &  $S_1\cdot S_3$  &  $S_1\cdot S_4$  &  $S_1\cdot S_5$  &  $S_2\cdot S_3$  &  $S_2\cdot S_4$  &  $S_2\cdot S_5$  &  $S_3\cdot S_4$  &  $S_3\cdot S_5$  &  $S_4\cdot S_5$  \\                                               
$\Braket{0,0,0,1/2 | S_i\cdot S_j |0,0,0,1/2}$  & $-3/4$ &  0   &  0  &  0  & 0  & 0  & 0  & $-3/4$ & 0 & 0 \\
$\Braket{0,0,0,1/2 | S_i\cdot S_j  |0,1,1,1/2}$  & 0 & 0  & 0 &  0  & 0  & 0  & 0 & 0  & $-\sqrt{3}/4$ &  $\sqrt{3}/4$  \\
$\Braket{0,1,1,1/2 | S_i\cdot S_j |0,1,1,1/2}$  & $-3/4$  & 0  & 0  &  0  & 0  & 0 & 0  & $1/4$ & $-1/2$ & $-1/2$ \\
$\Braket{0,1,1,3/2| S_i\cdot S_j |0,1,1,3/2}$  & $-3/4$  & 0  & 0 &  0  & 0  & 0 & 0  & $1/4$ & $1/4$ & $1/4$ \\
$\Braket{1,0,1,1/2| S_i\cdot S_j |1,0,1,1/2}$ & $1/4$ & 0 & 0  & $-1/2$ & 0 & 0 & $-1/2$ & $-3/4$  & 0 & 0 \\
$\Braket{1,0,1,1/2| S_i\cdot S_j |1,1,1,1/2}$ & 0 & $\sqrt{2}/4$ & $-\sqrt{2}/4$  &  0  & $\sqrt{2}/4$  & $-\sqrt{2}/4$            & 0 & 0  & $-\sqrt{2}/4$ & $\sqrt{2}/4$  \\
$\Braket{1,1,1,1/2| S_i\cdot S_j |1,1,1,1/2}$  & $1/4$  & $-1/4$  & $-1/4$  & $-1/4$  & $-1/4$ & $-1/4$  & $-1/4$ & $1/4$ & $-1/4$ & $-1/4$  \\
$\Braket{1,0,1,3/2| S_i\cdot S_j |1,0,1,3/2}$  & $1/4$ & 0  & 0 &  $1/4$  & 0  & 0  & $1/4$  & $-3/4$ & 0 & 0 \\
$\Braket{1,0,1,3/2| S_i\cdot S_j |1,1,1,3/2}$  & 0 & $\sqrt{2}/4$  & $-\sqrt{2}/4$  &  0 & $\sqrt{2}/4$  & $-\sqrt{2}/4$  & 0 & 0 & $\sqrt{2}/8$ & $-\sqrt{2}/8$ \\
$\Braket{1,0,1,3/2| S_i\cdot S_j |1,1,2,3/2}$ & 0 & 0 & 0 &  0 & 0  & 0  & 0 & 0 & $-\sqrt{5/32}$ & $\sqrt{5/32}$ \\
$\Braket{1,1,1,3/2| S_i\cdot S_j |1,1,1,3/2}$  & $1/4$ & $-1/4$  & $-1/4$  &  $1/8$ & $-1/4$ & $-1/4$ & $1/8$ & $1/4$     & $1/8$ & $1/8$  \\
$\Braket{1,1,1,3/2| S_i\cdot S_j |1,1,2,3/2}$ & 0 & 0  & 0 & $-\sqrt{5}/8$  & 0  & 0  & $-\sqrt{5}/8$ & 0 & $\sqrt{5}/8$ & $\sqrt{5}/8$ \\
$\Braket{1,1,2,3/2| S_i\cdot S_j |1,1,2,3/2}$ & $1/4$  & $1/4$  & $1/4$ & $-3/8$ & $1/4$ & $1/4$ & $-3/8$ & $1/4$     & $-3/8$ & $-3/8$  \\
$\Braket{1,1,2,5/2| S_i\cdot S_j |1,1,2,5/2}$ & $1/4$ & $1/4$ & $1/4$  & $1/4$  & $1/4$ & $1/4$  & $1/4$ & $1/4$ & $1/4$ & $1/4$ \\
$\Braket{0,0,0,1/2| S_i\cdot S_j |1,0,1,1/2}$ & 0 & 0 & 0  & $-\sqrt{3}/4$ & 0  & 0 & $\sqrt{3}/4$  & 0  & 0 & 0    \\
$\Braket{0,0,0,1/2| S_i\cdot S_j |1,1,1,1/2}$ & 0 & 0 & 0 & 0 & 0 & 0 & 0 & 0 & 0 & 0   \\
$\Braket{0,1,1,1/2| S_i\cdot S_j |1,0,1,1/2}$ & 0 & $1/4$ & $-1/4$ &  0  & $-1/4$ & $1/4$ & 0 & 0 & 0 &  0  \\
$\Braket{0,1,1,1/2| S_i\cdot S_j |1,1,1,1/2}$  & 0 & $-\sqrt{2}/4$ & $-\sqrt{2}/4$  & $\sqrt{2}/4$  & $\sqrt{2}/4$  & $\sqrt{2}/4$  & $-\sqrt{2}/4$   & 0     & 0 & 0    \\
$\Braket{0,1,1,3/2| S_i\cdot S_j |1,0,1,3/2}$  & 0  & $1/4$ & $-1/4$  &  0  & $-1/4$ & $1/4$ & 0 & 0 & 0 & 0   \\
$\Braket{0,1,1,3/2| S_i\cdot S_j |1,1,1,3/2}$   & 0  & $-\sqrt{2}/4$  & $-\sqrt{2}/4$   & $-\sqrt{2}/8$  & $\sqrt{2}/4$     & $\sqrt{2}/4$   & $\sqrt{2}/8$  & 0  & 0 & 0   \\
$\Braket{0,1,1,3/2| S_i\cdot S_j |1,1,2,3/2}$  & 0 & 0 & 0 & $-\sqrt{5/32}$  & 0  & 0  & $\sqrt{5/32}$  &  0  & 0 & 0  \\[2pt]
\hline\hline
\end{tabular*}
\end{table*}

In diquark models, the phenomenological Hamiltonian is usually parametrized as a sum of the masses of diquarks and quarks, and effective potentials composed of color-dependent interactions \cite{Maiani2005,Zhu2016,Shi2019,Kim2016}. In Ref.~\cite{Shi2020Tetra}, the flavor-dependent interactions are also introduced to describe the mass splits of the flavor multiplets. Following Ref.~\cite{Shi2020Tetra}, we write the effective Hamiltonian for the pentaquark systems as
\begin{align}
H= \sum_n M_n+m_c+V_c+V_F,   \label{Eq:Hamiltonian}
\end{align}
with $M_n$ being the effective mass of the $n$-th diquark and $m_c$ the mass of the charm antiquark. $V_C$ is the color-dependent interaction defined as
\begin{align}
V_C = 2\sum_{i>j} \left[ \alpha_{ij} \left( {\bm \lambda}^c_i \cdot {\bm \lambda}^c_j {\bm S}_i \cdot {\bm S}_j \right) + \frac{\beta}{m_im_j} \left( {\bm \lambda}^c_i \cdot {\bm \lambda}^c_j \right) \right],
\label{Eq:Hamiltonian_color}
\end{align}
where $\alpha_{ij}$ and $\beta$ are model parameters, and $m_i$ is the constituent quark mass. $V_F$ is the flavor-dependent interaction defined as
\begin{align}
V_{F} = 2\sum_{i>j} \left[ \left(\frac{\gamma}{m_{i}m_{j}}  {\bm \lambda}^f_{i} \cdot {\bm \lambda}^f_{j} {\bm S}_{i} \cdot {\bm S}_{j} \right) + \frac{\rho}{m_{i}m_{j}} \left( {\bm \lambda}^f_{i} \cdot {\bm \lambda}^f_{j} \right) \right],    \label{Eq:Hamiltonian_flavor}
\end{align}
where ${\bf \lambda}^f_{i}$ and ${\bf \lambda}^f_{j}$ are generators of the favor SU(3) group, and $\gamma$ and $\rho$ are model parameters. Note that the flavor-dependent interaction is inspired by the scalar meson exchanges and pseudoscalar meson exchanges stemming from the couplings of light quarks and chiral fields in the chiral quark model \cite{Huang2018}.

Considering that any hidden charm pentaquark state is color singlet, the color matrix elements can be calculated directly. The results are listed as follows:
\begin{align}
 \Braket{{\bm \lambda}^c_i \cdot {\bm \lambda}^c_j} = \left\{ \begin{array}{lll} -8/3, && \left(i, j {\rm ~ in ~ the ~ same ~ diquark}\right) \\[2pt]  -1/3, && \left( i ~{\rm or}~ j=5 \right)\\[2pt] -2/3. && \left({\rm others}\right) \end{array} \right.
 \label{Eq:color_matrix}
\end{align}

The spin wave function for a pentaquark state is constructed in Eq.~(\ref{eq:wf_spin}). For all possible spin states, the spin matrix elements are calculated and listed in Table~\ref{Tab:Spin_matrix}. 




\subsection{Mass formulas for the pentaquark states}

By use of Eq.~(\ref{Eq:color_matrix}) and Table~\ref{Tab:Spin_matrix}, the masses of hidden charm pentaquark states can be calculated by the effective Hamiltonian in Eq.~(\ref{Eq:Hamiltonian}).


The spin of the flavor singlet state could be $1/2$ or $3/2$. For the ${\bf{1}}_f$ state with spin $1/2$, its mass is determined by the following matrix:
\begin{align}
 \begin{pmatrix}
 M_{0, 0, 0, 1/2; \, {\bf{1}}_f }    & M'_{1, 2, 1/2;  \, {\bf{1}}_f } \\
 M'_{1, 2, 1/2; \,  {\bf{1}}_f }  &  M_{0, 1, 1, 1/2;  \, {\bf{1}}_f }                                                    
\end{pmatrix},
\label{Eq:mass_singlet_1_2}
\end{align}
with
\begin{align}
& M_{0,0,0,1/2;  \, {\bf{1}}_f } = \Braket{0,0,0,1/2;{\bf{1}}_f |H|0,0,0,1/2;{\bf{1}}_f },  \nonumber\\
& M_{0,1,1,1/2;  \, {\bf{1}}_f } = \Braket{0,1,1,1/2; {\bf{1}}_f |H|0,1,1,1/2; {\bf{1}}_f },  \nonumber\\
& M'_{1,2,1/2;  \, {\bf{1}}_f } = \Braket{0,0,0,1/2; {\bf{1}}_f |H|0,1,1,1/2;{\bf{1}}_f }.   \nonumber
\end{align}
Here the first four indices for each state are $S_{12}$, $S_{34}$, $S_{1234}$, and $S$, respectively, as defined in Eq.~(\ref{eq:wf_spin}). 
For the ${\bf{1}}_f$ state with spin $3/2$, its mass is given by
\begin{align}
M_{0,1,1,3/2;  \, {\bf{1}}_f} = \Braket{0,1,1,3/2;{\bf{1}}_f | H | 0,1,1,3/2; {\bf{1}}_f}.
\label{Eq:mass_singlet_3_2}
\end{align}

The spin of the flavor decuplet state could be $1/2$, $3/2$, or $5/2$. For the ${\bf{10}}_f$ state with spin $1/2$, its mass is determined by the following matrix:
\begin{align}
 \begin{pmatrix}
 M_{1,0,1,1/2; \, {\bf{10}}_f}    & M'_{1,2,1/2; \, {\bf{10}}_f} \\
 M'_{1,2,1/2; \, {\bf{10}}_f}  & M_{1,1,1,1/2; \, {\bf{10}}_f}                                                    
\end{pmatrix},
\label{Eq:mass_decuplet_1_2}
\end{align}
with
\begin{align}
& M_{1,0,1,1/2;  \, {\bf{10}}_f } = \Braket{1,0,1,1/2; {\bf{10}}_f | H | 1,0,1,1/2; {\bf{10}}_f},  \nonumber\\
& M_{1,1,1,1/2; \,  {\bf{10}}_f } = \Braket{1,1,1,1/2; {\bf{10}}_f | H | 1,1,1,1/2; {\bf{10}}_f},  \nonumber\\
& M'_{1,2,1/2;  \, {\bf{10}}_f } = \Braket{1,0,1,1/2; {\bf{10}}_f | H | 1,1,1,1/2; {\bf{10}}_f}.  \nonumber
\end{align}
For the ${\bf{10}}_f$ state with spin $3/2$, its mass is given by
\begin{align}
 \begin{pmatrix}
 M_{1,0,1,3/2; \, {\bf{10}}_f}    & M'_{1,2,3/2; \, {\bf{10}}_f}   & M'_{1,3,3/2; \, {\bf{10}}_f}\\
 M'_{1,2,3/2; \, {\bf{10}}_f}  & M_{1,1,1,3/2; \, {\bf{10}}_f}      & M'_{2,3,3/2; \, {\bf{10}}_f}\\   
 M'_{1,3,3/2; \, {\bf{10}}_f}  & M'_{2,3,3/2; \, {\bf{10}}_f}      & M_{1,1,2,3/2; \, {\bf{10}}_f}                                             
\end{pmatrix},
\label{Eq:mass_decuplet_3_2}
\end{align}
with
\begin{align}
&M_{1,0,1,3/2; \, {\bf{10}}_f} = \Braket{1,0,1,3/2;{\bf{10}}_f |H| 1,0,1,3/2;{\bf{10}}_f},  \nonumber\\
&M_{1,1,1,3/2; \, {\bf{10}}_f} = \Braket{1,1,1,3/2;{\bf{10}}_f |H| 1,1,1,3/2;{\bf{10}}_f},  \nonumber\\
&M_{1,1,2,3/2; \, {\bf{10}}_f} = \Braket{1,1,2,3/2;{\bf{10}}_f |H| 1,1,2,3/2;{\bf{10}}_f},  \nonumber\\
&M'_{1,2,3/2; \, {\bf{10}}_f} = \Braket{1,0,1,3/2;{\bf{10}}_f |H| 1,1,1,3/2;{\bf{10}}_f},  \nonumber\\
&M'_{1,3,3/2; \, {\bf{10}}_f} = \Braket{1,0,1,3/2;{\bf{10}}_f |H| 1,1,2,3/2;{\bf{10}}_f},  \nonumber\\
&M'_{2,3,3/2; \, {\bf{10}}_f} = \Braket{1,1,1,3/2;{\bf{10}}_f |H| 1,1,2,3/2;{\bf{10}}_f}.  \nonumber
\end{align}
For the ${\bf{10}}_f$ state with spin $5/2$, its mass is given by
\begin{align}
M_{1,1,2,5/2; \, {\bf{10}}_f} = \Braket{1,1,2,5/2;{\bf{10}}_f|H|1,1,2,5/2;{\bf{10}}_f}.
\label{Eq:mass_decuplet_5_2}
\end{align}

The spin of the flavor octet states could be $1/2$, $3/2$, or $5/2$. For the ${\bf{8}}^{\bf{S}}_f$ and ${\bf{8}}^{\bf{A}}_f$ states with spin $1/2$, its mass is determined by the following matrix:
\begin{align}
 \begin{pmatrix}
 M_{1,0,1,1/2; \, {\bf{8}}_f}    & M'_{1,2,1/2; \, {\bf{8}}_f}   & M'_{1,3,1/2; \, {\bf{8}}_f}  & M'_{1,4,1/2; \, {\bf{8}}_f}  \\
 M'_{1,2,1/2; \, {\bf{8}}_f}  & M_{1,1,1,1/2; \, {\bf{8}}_f}      & M'_{2,3,1/2; \, {\bf{8}}_f}   & M'_{2,4,1/2; \, {\bf{8}}_f}  \\   
 M'_{1,3,1/2; \, {\bf{8}}_f}  & M'_{2,3,1/2; \, {\bf{8}}_f}      & M_{0,0,0,1/2; \, {\bf{8}}_f}   & M'_{3,4,1/2; \, {\bf{8}}_f}  \\  
  M'_{1,4,1/2; \, {\bf{8}}_f}  & M'_{2,4,1/2; \, {\bf{8}}_f}      & M'_{3,4,1/2; \, {\bf{8}}_f}   & M_{0,1,1,1/2; \, {\bf{8}}_f}  \\                                        
\end{pmatrix},
\label{Eq:mass_octet_1_2}
\end{align}
with
\begin{align}
&M_{1,0,1,1/2; \, {\bf{8}}_f}=\Braket{1,0,1,1/2;{{\bf{8}}_f^{\bf{S}}} |H|1,0,1,1/2;{{\bf{8}}_f^{\bf{S}}}},  \nonumber\\
&M_{1,1,1,1/2; \, {\bf{8}}_f}=\Braket{1,1,1,1/2;{{\bf{8}}_f^{\bf{S}}} |H|1,1,1,1/2;{{\bf{8}}_f^{\bf{S}}}},  \nonumber\\
&M_{0,0,0,1/2; \, {\bf{8}}_f}=\Braket{0,0,0,1/2; {{\bf{8}}_f^{\bf{A}}} |H|0,0,0,1/2;{{\bf{8}}_f^{\bf{A}}}},  \nonumber\\
&M_{0,1,1,1/2; \, {\bf{8}}_f}=\Braket{0,1,1,1/2;{{\bf{8}}_f^{\bf{A}}}|H|0,1,1,1/2;{{\bf{8}}_f^{\bf{A}}}},  \nonumber\\
&M'_{1,2,1/2; \, {\bf{8}}_f}=\Braket{1,0,1,1/2;{{\bf{8}}_f^{\bf{S}}}|H|1,1,1,1/2;{{\bf{8}}_f^{\bf{S}}}},  \nonumber\\
&M'_{1,3,1/2; \, {\bf{8}}_f}=\Braket{1,0,1,1/2;{{\bf{8}}_f^{\bf{S}}}|H|0,0,0,1/2;{{\bf{8}}_f^{\bf{A}}}},  \nonumber\\
&M'_{1,4,1/2; \, {\bf{8}}_f}=\Braket{1,0,1,1/2;{{\bf{8}}_f^{\bf{S}}}|H|0,1,1,1/2;{{\bf{8}}_f^{\bf{A}}}},  \nonumber\\
&M'_{2,3,1/2; \, {\bf{8}}_f}=\Braket{1,1,1,1/2;{{\bf{8}}_f^{\bf{S}}}|H|0,0,0,1/2;{{\bf{8}}_f^{\bf{A}}}},  \nonumber\\
&M'_{2,4,1/2; \, {\bf{8}}_f}=\Braket{1,1,1,1/2;{{\bf{8}}_f^{\bf{S}}}|H|0,1,1,1/2;{{\bf{8}}_f^{\bf{A}}}},  \nonumber\\
&M'_{3,4,1/2; \, {\bf{8}}_f}=\Braket{0,0,0,1/2;{{\bf{8}}_f^{\bf{A}}}|H|0,1,1,1/2;{{\bf{8}}_f^{\bf{A}}}},  \nonumber
\end{align}
For the ${\bf{8}}_f$ state with spin $3/2$, its mass is given by
\begin{align}
 \begin{pmatrix}
 M_{1,0,1,3/2; \, {\bf{8}}_f}    & M'_{1,2,3/2; \, {\bf{8}}_f}   & M'_{1,3,3/2; \, {\bf{8}}_f}  & M'_{1,4,3/2; \, {\bf{8}}_f}  \\
 M'_{1,2,3/2; \, {\bf{8}}_f}  & M_{1,1,1,3/2; \, {\bf{8}}_f}      & M'_{2,3,3/2; \, {\bf{8}}_f}   & M'_{2,4,3/2; \, {\bf{8}}_f}  \\   
 M'_{1,3,3/2; \, {\bf{8}}_f}  & M'_{2,3,3/2; \, {\bf{8}}_f}      & M_{1,1,2,3/2; \, {\bf{8}}_f}   & M'_{3,4,3/2; \, {\bf{8}}_f}  \\  
  M'_{1,4,3/2; \, {\bf{8}}_f}  & M'_{2,4,3/2; \, {\bf{8}}_f}      & M'_{3,4,3/2; \, {\bf{8}}_f}   & M_{0,1,1,3/2; \, {\bf{8}}_f}  \\                                        
\end{pmatrix},
\label{Eq:mass_octet_3_2}
\end{align}
with
\begin{align}
&M_{1,0,1,3/2; \, {\bf{8}}_f} = \Braket{1,0,1,3/2; {{\bf{8}}_f^{\bf{S}}}|H|1,0,1,3/2; {{\bf{8}}_f^{\bf{S}}} },  \nonumber\\
&M_{1,1,1,3/2; \, {\bf{8}}_f} = \Braket{1,1,1,3/2; {{\bf{8}}_f^{\bf{S}}}|H|1,1,1,3/2; {{\bf{8}}_f^{\bf{S}}} },  \nonumber\\
&M_{1,1,2,3/2; \, {\bf{8}}_f} = \Braket{1,1,2,3/2; {{\bf{8}}_f^{\bf{S}}}|H|1,1,2,3/2; {{\bf{8}}_f^{\bf{S}}} },  \nonumber\\
&M_{0,1,1,3/2; \, {\bf{8}}_f} = \Braket{0,1,1,3/2; {{\bf{8}}_f^{\bf{A}}}|H|0,1,1,3/2; {{\bf{8}}_f^{\bf{A}}} },  \nonumber\\
&M'_{1,2,3/2; \, {\bf{8}}_f} = \Braket{1,0,1,3/2; {{\bf{8}}_f^{\bf{S}}}|H|1,1,1,3/2; {{\bf{8}}_f^{\bf{S}}} },  \nonumber\\
&M'_{1,3,3/2; \, {\bf{8}}_f} = \Braket{1,0,1,3/2; {{\bf{8}}_f^{\bf{S}}}|H|1,1,2,3/2; {{\bf{8}}_f^{\bf{S}}} },  \nonumber\\
&M'_{1,4,3/2; \, {\bf{8}}_f} = \Braket{1,0,1,3/2; {{\bf{8}}_f^{\bf{S}}}|H|0,1,1,3/2; {{\bf{8}}_f^{\bf{A}}} },  \nonumber\\
&M'_{2,3,3/2; \, {\bf{8}}_f} = \Braket{1,1,1,3/2; {{\bf{8}}_f^{\bf{S}}}|H|1,1,2,3/2; {{\bf{8}}_f^{\bf{S}}} },  \nonumber\\
&M'_{2,4,3/2; \, {\bf{8}}_f} = \Braket{1,1,1,3/2; {{\bf{8}}_f^{\bf{S}}}|H|0,1,1,3/2; {{\bf{8}}_f^{\bf{A}}} },  \nonumber\\
&M'_{3,4,3/2; \, {\bf{8}}_f} = \Braket{1,1,2,3/2; {{\bf{8}}_f^{\bf{S}}}|H|0,1,1,3/2; {{\bf{8}}_f^{\bf{A}}} }.  \nonumber
\end{align}
For the ${\bf{8}}_f$ state with spin $5/2$, its mass is given by
\begin{align}
M_{1,1,2,5/2; \, {\bf{8}}_f} = \Braket{1,1,2,5/2;{{\bf{8}}_f^{\bf{S}}}|H|1,1,2,5/2;{{\bf{8}}_f^{\bf{S}}}}.
\label{Eq:mass_octet_5_2}
\end{align}


\subsection{The model parameters} \label{Sect:parameter}

\begin{table}[b]
\caption{\label{Tab:parameters} Model Parameters. The diquark masses $M_{qq}$, $M_{qc}$, $M_{sc}$, and the coefficients $\alpha_{qq}$, $\alpha_{qs}$, $\alpha_{qc}$, $\alpha_{sc}$, $\alpha_{cc}$, $\alpha_{ss}$ are in MeV. The coefficients $\beta$, $\rho$, and $\gamma$ are in fm$^{-3}$.}
\renewcommand{\arraystretch}{1.3}
\begin{tabular*}{\columnwidth}{@{\extracolsep\fill}cccccc}
\hline\hline                                           
 $M_{qq}$  & $M_{qc}$ & $M_{qs}$ & $\beta$ & $\rho$ & $\gamma$  \\
 $1024$     &  $2059$   &  $1098$  &  $0.41$  & $0.11$  & $0.03$  \\ 
 \hline
 $\alpha_{qq}$ & $\alpha_{qs}$  & $\alpha_{qc}$ & $\alpha_{sc}$ & $\alpha_{ss}$ & $\alpha_{cc}$  \\
 $-28.43$       &  $-21.70$        &  $-8.06$       &  $-8.98$        &  $-23.24$      &  $-10.59$  \\
\hline\hline
\end{tabular*}
\end{table}

As shown in Eq.~(\ref{Eq:Hamiltonian}), the diquark masses $M_{qq}$, $M_{qs}$, $M_{ss}$, $M_{qc}$, and $M_{sc}$ are model parameters. We use the following relation to get $M_{ss}$ from $M_{qs}$ and $M_{qq}$ with $q=u,d$ \cite{JAFFE2005}:
\begin{align}
M_{ss}=2M_{qs}-M_{qq}.
\label{Eq:mass_diquark_assumption}
\end{align}
The parameters of these diquark masses and also the other model parameters including $\alpha_{qq}$, $\alpha_{qs}$, $\alpha_{qc}$, $\alpha_{sc}$, $\alpha_{ss}$, $\alpha_{cc}$, $\beta$, $\gamma$, and $\rho$ need to be fixed before the calculation of the masses of the considered hidden charm pentaquark states.

Following Refs.~\cite{Shi2019,Shi2020Tetra}, we use the masses of the heavy baryons $\Lambda_c$, $\Xi_c$, $\Omega_c^0$, $\Sigma_c$, $\Xi_c'$, $\Sigma_c^*$, and $\Xi^*_c$, which are attributed to diquark-quark configurations, to fix the parameters $M_{qq}$, $M_{qs}$, $\alpha_{qq}$, $\alpha_{qs}$, $\alpha_{qc}$, $\alpha_{sc}$, $\alpha_{ss}$, and $\beta$. In addition, the $X(3872)$, $Z_c(3900)$, and $\chi_{c2}(3930)$ are attributed as $[qc][\bar{q}\bar{c}]$ states with $I^GJ^{PC}$ being $0^+1^{++}$, $1^+1^{+-}$, and $0^+2^{++}$, respectively, to fixed the diquarks mass $M_{qc}$ and the coefficients $\rho$ and $\gamma$. 

The masses of $u$, $d$, $s$, and $c$ quarks are taken to be $m_{u(d)}=313$ MeV, $m_s=470$ MeV, and $m_c=1650$ MeV \cite{Huang2018,Shi2019}. 

The diquark mass $M_{sc}$ is fixed by the mass of the $X(4350)$ \cite{Shi2020Tetra}. In model $\rm{\uppercase\expandafter{\romannumeral1}}$, the spin of $X(4350)$ is supposed to be 0, which gives $M_{sc}=2252$ MeV, and in model $\rm{\uppercase\expandafter{\romannumeral2}}$, the spin of $X(4350)$ is supposed to be 2, which gives $M_{sc}=2205$ MeV. 

The values of the parameters $M_{qq}$, $M_{qs}$, $M_{qc}$, $\alpha_{qq}$, $\alpha_{qs}$, $\alpha_{qc}$, $\alpha_{sc}$, $\alpha_{ss}$, $\alpha_{cc}$, $\beta$, $\rho$, and $\gamma$ are listed in Table~\ref{Tab:parameters}. 

Note that the diquark masses include the effects that are not accounted for by the interactions given in the effective model Hamiltonian. In this sense, they are highly model dependent. As explained above, in the present work and also our previous works of Refs.~\cite{Shi2019,Shi2020Tetra}, the light diquark mass $M_{qq}$ is fixed by the masses of the heavy baryons $\Lambda_c$, $\Sigma_c$, and $\Sigma_c^\ast$. In Ref.~\cite{Karliner:2015ema}, a smaller light diquark mass was used. There, the value of $M_{qq}$ was determined by the masses of the light baryons $\Lambda$, $\Sigma$, and $\Sigma^\ast$.

In principle the color-magnetic interaction has explicitly $1/(m_im_j)$ dependence. In the present work, we simply follow Refs.~\cite{Maiani2015,Anisovich2015,Lebed2015,Li2015,Wang2016,Chen2016,Ali2016,Ghosh2017,Ali:2019npk,Semenova2019} to parametrize the color-spin dependent interaction as $\alpha_{ij}  {\bm \lambda}^c_i \cdot {\bm \lambda}^c_j {\bm S}_i \cdot {\bm S}_j $, as shown in Eq.~(\ref{Eq:Hamiltonian_color}). One may define $\alpha'_{ij}=\alpha_{ij}m_im_j$, which can be compared directly to the parameter in the traditional color-magnetic interaction $\frac{\alpha'_{ij}}{m_im_j} {\bm \lambda}^c_i \cdot {\bm \lambda}^c_j {\bm S}_i \cdot {\bm S}_j $. By using the values of $\alpha_{ij}$ in Table~\ref{Tab:parameters}, one would get $|\alpha'_{qq}|<|\alpha'_{qs}|<|\alpha'_{ss}|$, and $|\alpha'_{qc}|<|\alpha'_{sc}|<|\alpha'_{cc}|$, which are consistent with those observed in microscopic quark model calculations \cite{Huang2018,WWL2011}.

\section{Numerical results}\label{Sect:Numerical_result}

\begin{table}[tb]
\caption{\label{mass_flavor_qqqcc} Masses (in MeV) of $(qq)(qc){\bar{c}}$ ($q=u,d$) configurations. The first column denotes the SU(3) flavor multiplet of the pentaquark state. The second and third columns denote, respectively, the isospin spin-parity $IJ^{P}$ and mass of the $(qq)(qc){\bar{c}}$ state predicted in the present work. The last two columns denote the particle name and mass cited in Review of Particle Physics (RPP) \cite{PDG2020}.}
\renewcommand{\arraystretch}{1.3}
\begin{tabular*}{\columnwidth}{@{\extracolsep\fill}ccccc}
\hline\hline                                           
Multiplet  & $\left(IJ^{P}\right)_{\rm th.}$ & $M_{\rm th.}$ &  Particle    &  $M_{\rm PDG}$  \\[2pt]                                               
\hline
${\bf{8}}_f$ & $\frac{1}{2} \frac{1}{2}^{-}$  & $4238$  &  &    \\[2pt]
                 & $\frac{1}{2} \frac{1}{2}^{-}$   & $4274$  &   &    \\[2pt]
                 & $\frac{1}{2} \frac{1}{2}^{-}$   & $4370$  &   &    \\[2pt]
                 & $\frac{1}{2} \frac{1}{2}^{-}$   & $4406$  &  &    \\[2pt]
                 & $\frac{1}{2} \frac{3}{2}^{-}$  & $4284$  & $P_c(4312)^+$     &  $4311.9\!\pm\!0.7^{+6.8}_{-0.6}$  \\[2pt]
                & $\frac{1}{2} \frac{3}{2}^{-}$   & $4379$   &   &  \\[2pt]
                & $\frac{1}{2} \frac{3}{2}^{-}$    & $4413$   &    &  \\[2pt]
                & $\frac{1}{2} \frac{3}{2}^{-}$    & $4436$  & $P_c(4440)^+$    &  $4440.3\!\pm\!1.3^{+4.1}_{-4.7}$ \\[2pt]
                & $\frac{1}{2} \frac{5}{2}^{-}$   & $4451$    & $P_c(4457)^+$    &  $4457.3\!\pm\!0.6^{+4.1}_{-1.7}$ \\[2pt]
\hline
${\bf{10}}_f$ & $\frac{3}{2} \frac{1}{2}^{-}$   &  $4464$   &      &   \\[2pt]
                  & $\frac{3}{2} \frac{1}{2}^{-}$    &  $4514$   &   &  \\[2pt]
                  & $\frac{3}{2} \frac{3}{2}^{-}$   &  $4473$   &   &   \\[2pt]
                  & $\frac{3}{2} \frac{3}{2}^{-}$   &  $4522$   &    &  \\[2pt]
                  & $\frac{3}{2} \frac{3}{2}^{-}$   &  $4549$   &   &  \\[2pt]
                  & $\frac{3}{2} \frac{5}{2}^{-}$   &  $4563$   &   &  \\[2pt]
\hline\hline
\end{tabular*}
\end{table}

\begin{table}[tb]
\caption{\label{mass_flavor_qqscc} Masses (in MeV) of $(qq)(qc){\bar{c}}$ ($q=u,d,s$) configurations with strangeness $S=-1$. The first two columns denote the SU(3) flavor multiplet and the isospin spin-parity $IJ^{P}$ of the pentaquark state predicted in the present work. The third and fourth columns denote the masses calculated in model I and model II, respectively. The last two columns denote the particle name and mass cited in Review of Particle Physics (RPP) \cite{PDG2020}. In the fifth column, the symbol ``?'' means that the corresponding state has more than one possible assignment.}
\renewcommand{\arraystretch}{1.3}
\begin{tabular*}{\columnwidth}{@{\extracolsep\fill}cccccc}
\hline\hline                                           
Multiplet        &  $\left(IJ^{P}\right)_{\rm th.}$          & Model I   & Model II  & Particle &  $M_{\rm PDG}$  \\[2pt]                                               
\hline
${\bf{1}}_f$   & $0 \frac{1}{2}^{-}$   &  $4357$   &  $4342$      \\[2pt]
                   & $0 \frac{1}{2}^{-}$   &  $4396$   &  $4380$     \\[2pt]
                   & $0 \frac{3}{2}^{-}$  &  $4405$   &  $4390$     \\[2pt]
\hline
${\bf{8}}_f$  & $0 \frac{1}{2}^{-}$    & $4446$   &  $4415$   & $P_{cs}(4459)$?  &  $4458.8\!\pm\!2.9^{+4.7}_{-1.1}$ \\[2pt]
                  & $0 \frac{1}{2}^{-}$    & $4485$   &  $4454$  &  $P_{cs}(4459)$?  &  $4458.8\!\pm\!2.9^{+4.7}_{-1.1}$      \\[2pt]
                  & $0 \frac{1}{2}^{-}$    & $4510$    &  $4510$      \\[2pt]
                  & $0 \frac{1}{2}^{-}$    & $4542$   &  $4542$     \\[2pt]
                  & $1 \frac{1}{2}^{-}$    &  $4432$   &  $4432$     \\[2pt]
                  & $1 \frac{1}{2}^{-}$    &  $4469$   &  $4469$  &   &   \\[2pt]
                  & $1 \frac{1}{2}^{-}$    &  $4593$   &  $4562$     \\[2pt]
                  & $1 \frac{1}{2}^{-}$    &  $4628$   &  $4597$     \\[2pt]
                  & $0 \frac{3}{2}^{-}$   & $4494$    &  $4463$  &  $P_{cs}(4459)$?  &  $4458.8\!\pm\!2.9^{+4.7}_{-1.1}$    \\[2pt]
                  & $0 \frac{3}{2}^{-}$   & $4519$    &   $4519$     \\[2pt]
                  & $0 \frac{3}{2}^{-}$   & $4550$   &  $4550$     \\[2pt]
                  & $0 \frac{3}{2}^{-}$   & $4572$   &  $4572$     \\[2pt]
                  & $1 \frac{3}{2}^{-}$   & $4478$   &  $4478$     \\[2pt]
                  & $1 \frac{3}{2}^{-}$   & $4602$   & $4571$     \\[2pt]
                  & $1 \frac{3}{2}^{-}$   & $4636$    & $4605$     \\[2pt]
                  & $1 \frac{3}{2}^{-}$   & $4657$    & $4626$     \\[2pt]
                  & $0 \frac{5}{2}^{-}$  & $4587$     & $4587$     \\[2pt]
                 & $1 \frac{5}{2}^{-}$    & $4672$    &  $4641$     \\[2pt] 
\hline 
${\bf{10}}_f$   & $1 \frac{1}{2}^{-}$   & $4625$  & $4610$     \\[2pt]
                    & $1 \frac{1}{2}^{-}$    & $4668$  & $4652$     \\[2pt]
                    & $1 \frac{3}{2}^{-}$   & $4635$   & $4619$     \\[2pt]
                    & $1 \frac{3}{2}^{-}$   & $4676$   & $4661$    \\[2pt]
                    & $1 \frac{3}{2}^{-}$   & $4702$   & $4686$     \\[2pt]
                    & $1 \frac{5}{2}^{-}$   & $4717$    & $4701$     \\[2pt]
\hline\hline
\end{tabular*}
\end{table}

\begin{table}[tb]
\caption{\label{mass_flavor_qsscc} Masses (in MeV) of $(qq)(qc){\bar{c}}$ ($q=u,d,s$) configurations with strangeness $S=-2$. The first two columns denote the SU(3) flavor multiplet and the isospin spin-parity $IJ^{P}$ of the pentaquark state predicted in the present work. The last two columns denote the masses calculated in model I and model II, respectively.}
\renewcommand{\arraystretch}{1.3}
\begin{tabular*}{\columnwidth}{@{\extracolsep\fill}cccc}
\hline\hline                                           
Multiplet        & $\left(IJ^{P}\right)_{\rm th.}$          & Model I   & Model II    \\[2pt]                                               
\hline
${\bf{8}}_f$   & $\frac{1}{2} \frac{1}{2}^{-}$   &  $4645$  &  $4598$    \\[2pt]
                   & $\frac{1}{2} \frac{1}{2}^{-}$    &  $4682$  & $4639$    \\[2pt]
                   & $\frac{1}{2} \frac{1}{2}^{-}$    &  $4689$  & $4670$    \\[2pt]
                   & $\frac{1}{2} \frac{1}{2}^{-}$    &  $4717$   &  $4702$    \\[2pt]
                  & $\frac{1}{2} \frac{3}{2}^{-}$    &  $4691$   &  $4649$     \\[2pt]
                  & $\frac{1}{2} \frac{3}{2}^{-}$    &  $4698$  & $4679$     \\[2pt]
                  & $\frac{1}{2} \frac{3}{2}^{-}$    &  $4726$  & $4710$      \\[2pt]
                  & $\frac{1}{2} \frac{3}{2}^{-}$   &  $4747$  &  $4731$     \\[2pt]
                  & $\frac{1}{2} \frac{5}{2}^{-}$    &  $4762$  & $4746$     \\[2pt]
\hline         
${\bf{10}}_f$  & $\frac{1}{2} \frac{1}{2}^{-}$   & $4780$   & $4749$     \\[2pt]
                    & $\frac{1}{2} \frac{1}{2}^{-}$   & $4820$   & $4789$    \\[2pt]        
                    & $\frac{1}{2} \frac{3}{2}^{-}$   & $4790$   & $4759$     \\[2pt]
                   & $\frac{1}{2} \frac{3}{2}^{-}$    & $4828$   & $4797$     \\[2pt]
                  & $\frac{1}{2} \frac{3}{2}^{-}$     & $4854$   & $4823$    \\[2pt]
                 & $\frac{1}{2} \frac{5}{2}^{-}$     &  $4869$   & $4838$     \\[2pt]
\hline\hline
\end{tabular*}
\end{table}

\begin{table}[htb]
\caption{\label{mass_flavor_ssscc} Masses (in MeV) of $(ss)(sc){\bar{c}}$ configurations. The first two columns denote the SU(3) flavor multiplet and the isospin spin-parity $IJ^{P}$ of the pentaquark state predicted in the present work. The last two columns denote the masses calculated in model I and model II, respectively.}
\renewcommand{\arraystretch}{1.3}
\begin{tabular*}{\columnwidth}{@{\extracolsep\fill}cccc}
\hline\hline                                           
Multiplet       & $\left(IJ^{P}\right)_{\rm th.}$          & Model I   & Model II    \\[2pt]                                               
\hline
${\bf{10}}_f$   & $0 \frac{1}{2}^{-}$   &  $4929$     &  $4883$     \\[2pt]
                    & $0 \frac{1}{2}^{-}$    &  $4969$     &  $4923$     \\[2pt]
                    & $0 \frac{3}{2}^{-}$    &  $4939$    &  $4892$     \\[2pt]
                    & $0 \frac{3}{2}^{-}$    &  $4978$     &  $4932$    \\[2pt]
                    & $0 \frac{3}{2}^{-}$   &  $5004$      &  $4958$    \\[2pt]
                   &  $0 \frac{5}{2}^{-}$   &  $5020$      &  $4973$     \\[2pt]
\hline\hline
\end{tabular*}
\end{table}

With the values of model parameters listed in Table~\ref{Tab:parameters}, the masses of hidden charm pentaquark configurations composed of diquark-diquark-antiquark with orbit angular momentum $L=0$ can be computed from Eqs.~(\ref{Eq:mass_decuplet_1_2})$-$(\ref{Eq:mass_octet_5_2}). The numerical results for $(qq)(qc){\bar{c}}$ ($q=u,d$) are listed in Table~\ref{mass_flavor_qqqcc}. There, the first column denotes the SU(3) flavor multiplet of the pentaquark state. The second and third columns denote the isospin $I$, spin-parity $J^{P}$, and mass of the $(qq)(qc){\bar{c}}$ state predicted in the present work. The last two columns denote the particle name and mass cited in Review of Particle Physics (RPP) \cite{PDG2020}. 
 
From Table~\ref{mass_flavor_qqqcc} one sees that the experimentally observed $P_c(4312)^+$ and $P_c(4440)$ states can be well accommodated as $(qq)(qc){\bar{c}}$ states with $IJ^{P}=1/2\left(3/2^-\right)$ in the present diquark model. For $P_{c}(4457)$, its mass is very close to our calculated masses for $(qq)(qc){\bar{c}}$ states with $IJ^{P}=1/2\left(5/2^-\right)$ and $3/2\left(1/2^-\right)$. However, due to the fact that the $P_{c}(4457)$ can decay to $J/\psi p$ \cite{PDG2020}, its isospin should be $1/2$ and the possibility $IJ^{P}=3/2\left(1/2^-\right)$ is ruled out. Note that for all the $P_c$ states, the experimental quantum numbers are still not yet known, and our calculation serve as possible theoretical suggestions.

Our numerical results for $(qq)(qc){\bar{c}}$ ($q=u,d,s$) configurations with strangeness $S=-1$ are listed in Table~\ref{mass_flavor_qqscc}. There, the first two columns denote the SU(3) flavor multiplet and the isospin spin-parity $IJ^{P}$ of each pentaquark state predicted in the present work. The third and fourth columns denote the masses calculated in model I and model II, respectively. The last two columns denote the particle name and mass cited in Review of Particle Physics (RPP) \cite{PDG2020}. In the fifth column, the symbol ``?'' means that the corresponding state has more than one possible assignment. Actually, one sees from Table~\ref{mass_flavor_qqscc} that in model I, the experimentally observed $P_{cs}(4459)$ can be assigned to a $(qq)(qc){\bar{c}}$ ($q=u,d,s$) state with strangeness $S=-1$ and isospin spin-parity $IJ^{P}=0\left(1/2^-\right)$, while in model II, the $P_{cs}(4459)$ particle can be assigned to one of the $(qq)(qc){\bar{c}}$ ($q=u,d,s$) states with $S=-1$ and $IJ^{P}=0\left(1/2^-\right)$ or $0\left(3/2^-\right)$, respectively. Note that the $P_{cs}(4459)$ has isospin $0$ as it can decay to $\Lambda J/\psi$. We mention that the model I and model II differ in such a way that the model parameters are fixed by assuming that the spin of $X(4350)$ is 0 in model I and 2 in model II, respectively. Thus, the experimental information for the quantum numbers of $P_{cs}(4459)$ will help to distinguish the model I and model II, and further to understand the quantum numbers of $X(4350)$, and vice versa.

The calculated masses for $(qq)(qc){\bar{c}}$ ($q=u,d,s$) configurations with strangeness $S=-2$ and $-3$ are listed in Tables~\ref{mass_flavor_qsscc} and \ref{mass_flavor_ssscc}, respectively. One sees that generally, the states in flavor ${\bf{10}}_f$ multiplet are heavier than those in flavor ${\bf{8}}_f$ multiplet, and the states with $S=-3$ are heavier than those with $S=-2$. As the experimental information for these states are still not yet available, our theoretical results serve as predictions. Hope that these states can be searched in experiments in the near future.

We mention that the $P_c$ and $P_{cs}$ states have also been investigated in literature where no flavor-dependent interactions were considered compared with the Hamiltonian employed in the present work. For the $P_c(4312)^+$ state, our present work and the works of Refs.~\cite{Ali:2019npk,Ali:2019clg} suggest an assignment of its spin-parity $J^P=3/2^-$, while in Ref.~\cite{Semenova2019}, it was suggested to have $J^P=1/2^-$. For the $P_c(4440)^+$ state, our present work suggests $J^P=3/2^-$, Refs.~\cite{Ali:2019npk,Ali:2019clg} suggested $J^P=3/2^+$, and Ref.~\cite{Semenova2019} suggested $J^P=1/2^-$, respectively. For the $P_c(4457)^+$ state, our present work suggests $J^P=5/2^-$, Refs.~\cite{Ali:2019npk,Ali:2019clg} suggested $J^P=5/2^+$, and Ref.~\cite{Semenova2019} suggested $J^P=3/2^-$, respectively. For the $P_{cs}(4459)$ state, our present work suggests its quantum numbers to be $J^P=1/2^-$ or $3/2^-$. In Refs.~\cite{Ali:2019npk,Ali:2019clg}, a pentaquark $\left(\left[cs\right]_{\bar 3}\left[qq'\right]_{\bar 3}{\bar c}_{\bar 3}\right)$ $(q'=u,d)$ was predicted with quantum numbers $J^P=1/2^-$ or $3/2^-$ and mass $M=4488\pm 25$ MeV which is very close to the experimental value of $P_{cs}(4459)$. More experimental information for these states are called on to distinguish these theoretical results.

\section{The decay mode of hidden charm pentaquark states} \label{Sec:decay}

\begin{table*}[htb]
\caption{\label{Tab:Spin_rearrange} Recoupling coefficients of the spin wave functions for pentaquark states. The spin wave function $\Ket{S_{12},S_{34},S_{1234},S}$ in the first column is defined in Eq.~(\ref{eq:wf_spin}). The rearranged spin wave function in the first row is denoted by $\Ket{S_{123},S_{4{\bar 5}},S}$. }
\renewcommand{\arraystretch}{1.3}
\begin{tabular*}{\textwidth}{@{\extracolsep\fill}lccccccc}
\hline\hline                                           
  &  $\Ket{1/2, 0, 1/2} $  &  $\Ket{1/2, 1, 1/2}$  &  $\Ket{1/2, 1, 3/2}$  &  $\Ket{3/2, 0, 3/2}$  &  $\Ket{3/2, 1, 1/2}$ &  $\Ket{3/2, 1, 3/2}$  &  $\Ket{3/2, 1, 5/2}$   \\ 
$\Ket{0, 0, 0, 1/2}$  & $-1/2$ & $\sqrt{3}/2$   &  0  &  0  & 0  & 0 & 0  \\
$\Ket{0, 1, 1, 1/2}$  & $\sqrt{3}/2$ & $1/2$  & 0 &  0  & 0  & 0 & 0   \\
$\Ket{0, 1, 1, 3/2}$  & 0  & 0  & 1  &  0  & 0  & 0 & 0 \\
$\Ket{1, 0, 1, 1/2}$  & $-1/2$  & $-1/\sqrt{12}$  & 0 &  0  & $\sqrt{2/3}$  & 0 & 0 \\
$\Ket{1, 0, 1, 3/2}$ & 0 & 0 & $-1/\sqrt{3}$  & $-1/2$ & 0 & $\sqrt{5/12}$ & 0 \\
$\Ket{1, 1, 1, 1/2}$ & $1/\sqrt{2}$ & $1/\sqrt{6}$ & 0  &  0  & $1/\sqrt{3}$  & 0 & 0\\
$\Ket{1, 1, 1, 3/2}$  & 0  & 0  & $\sqrt{2/3}$  & $-1/\sqrt{8}$  & 0 & $\sqrt{5/24}$  & 0  \\
$\Ket{1, 1, 2, 3/2}$  & 0 & 0  & 0  & $\sqrt{5/8}$  & 0  & $\sqrt{3/8}$ & 0  \\
$\Ket{1, 1, 2, 5/2}$  & 0 & 0  & 0  & 0  & 0  & 0 & 1  \\[2pt]
\hline\hline
\end{tabular*}
\end{table*}

In the diquark model, the decay of a pentaquark state can be described as a tunneling process, i.e. two quarks tunnel from a diquark to another diquark and an antiquark and then they rearrange to two hadrons \cite{Maiani:2017kyi}. In a semi-classical approximation, the tunneling amplitude can be expressed as ${\cal{A}} \sim e^{-r\sqrt{2m\Delta E}}$, with $r$ being the distance of diquarks, $m$ the mass of a tunneling quark, and $\Delta E$ the depth of potential barrier that a quark inside a diquark needs to tunnel through \cite{Maiani:2017kyi,Shi2019}. Due to the heavy quark symmetry \cite{Isgur:1991wq,Ali:2019clg}, a hidden charm pentaquark state $(q'q')(q'c)\bar{c}$ $(q'=u,d,s)$ decays in such a way that the quarks inside the diquark $(q'c)$ tunnel to the diquark $(q'q')$ and antiquark $\bar{c}$ to form a baryon and a meson, i.e. $(q'q')c$ and $q'\bar{c}$, or $(q'q')q'$ and $c\bar{c}$. Here we make a rough discussion about the possible decay channels of hidden charm pentaquark states via the rearrangements of their color and spin wave functions instead of performing a real calculation of their partial decay widths. 

The color wave function of a hidden charm pentaquark state $(q'q')(q'c)\bar{c}$ $(q'=u,d,s)$ depicted in Eq.~(\ref{Eq:color_wave_function}) reads
\begin{align}
\psi_c = \frac{1}{\sqrt{6}} \epsilon_{ijk} \left(\frac{1}{\sqrt{2}}\epsilon^{imn}q'_{m}q'_{n}\right) \left(\frac{1}{\sqrt{2}}\epsilon^{jhl}q'_{h}c_{l}\right) \bar{c}^k,
\end{align}
with $i$, $j$, $k$, $m$, $n$, $h$, and $l$ being the color indices. It can be rearranged as 
\begin{align}
\psi_c = & - \frac{1}{2\sqrt{6}} \left[\left(\epsilon^{imn}q'_{m}q'_{n}q'_{i}\right) \left(c_k\bar{c}^k\right) - \left(\epsilon^{imn}q'_{m}q'_{n}c_{i}\right) \left(q'_k\bar{c}^k\right) \right]  \nonumber  \\[3pt]
          = & - \frac{\sqrt{3}}{2} \left\{\left[\left(q'q'\right)_{\bar{\bf 3}}q'\right]_{\bf 1_c}\left(c\bar{c}\right)_{\bf 1_c} - \left[\left(q'q'\right)_{\bar{\bf 3}}c\right]_{\bf 1_c}\left(q'\bar{c}\right)_{\bf 1_c}\right\},       \label{Eq:color_rearrange}
\end{align}
where the notations $\left[\left(q'q'\right)_{\bar{\bf 3}}q'\right]_{\bf 1_c}$ and $\left[\left(q'q'\right)_{\bar{\bf 3}}c\right]_{\bf 1_c}$ represent baryons, and $\left(c\bar{c}\right)_{\bf 1_c}$ and $\left(q'\bar{c}\right)_{\bf 1_c}$ represent mesons.

The spin wave function of a pentaquark state given in Eq.~(\ref{eq:wf_spin}) can be rearranged as 
\begin{align}
\Ket{S_{12},S_{34},S_{1234},S} \equiv & \left\{\left[\left(s_1s_2\right)_{S_{12}} \left(s_3s_4\right)_{S_{34}} \right]_{S_{1234}} s_{\bar 5}\right\}_S  \nonumber  \\[3pt]
    = & \sum_{S_{123},S_{4{\bar 5}}} (-1)^{S_{12}+S_{123}+S_{1234}+S} \hat{S}_{123}\hat{S}_{34}\hat{S}_{1234}\hat{S}_{4{\bar 5}}  \nonumber  \\[3pt]
 & \times \begin{Bmatrix}
           S_{12}  & s_3  & S_{123}  \\  s_4  &  S_{1234}  &  S_{34}
    \end{Bmatrix}
    \begin{Bmatrix}
      S_{123}  & s_4  & S_{1234}  \\  s_{\bar 5}  &  S  &  S_{4{\bar 5}}
    \end{Bmatrix}    \nonumber    \\[3pt]
 & \times \left[\left(s_1s_2s_3\right)_{S_{123}}\left(s_4s_{\bar 5}\right)_{S_{4{\bar 5}}}\right]_S,    \label{Eq:spin_rearrange}
\end{align}
where $S_{123}$ and $S_{4{\bar 5}}$ are the spins of the three-quark and quark-antiquark clusters, respectively. The notation $\hat{S}$ is defined as $\sqrt{2S+1}$. The rearranged spin wave functions given by Eq.~(\ref{Eq:spin_rearrange}) are listed in Table~\ref{Tab:Spin_rearrange}, where the notation $\Ket{S_{123},S_{4{\bar 5}},S}$ is used in the first row and the notation $\Ket{S_{12},S_{34},S_{1234},S}$ defined in Eq.~(\ref{eq:wf_spin}) is used in the first column.

According to the rearrangements of the color wave function in Eq.~(\ref{Eq:color_rearrange}) and the spin wave function in Table~\ref{Tab:Spin_rearrange}, the $P_c$ states with $I(J^P)=1/2(3/2^-)$ can rearrange to $NJ/\psi$, $\Lambda_c \bar{D}^\ast$, $\Sigma_c \bar{D}^\ast$, $\Sigma_c^\ast \bar{D}$, and $\Sigma_c^\ast \bar{D}^\ast$ channels. As its predicted mass is below the thresholds of $\Lambda_c \bar{D}^\ast$, $\Sigma_c \bar{D}^\ast$, $\Sigma_c^\ast \bar{D}$, and $\Sigma_c^\ast \bar{D}^\ast$, the $P_c(4312)$ is expected to decay to the $NJ/\psi$ channel via $S$ wave. Note that the experimental mass of $P_c(4312)$ is also above the threshold of $\Lambda_c \bar{D}^\ast$. If its quantum numbers are finally confirmed to be $I(J^P)=1/2(3/2^-)$, it may decay to the $\Lambda_c \bar{D}^\ast$ channel via $S$-wave, too. The $P_c(4440)$ is predicted to have the same quantum numbers as the $P_c(4312)$, but it has a higher mass which is above the threshold of $\Sigma_c^\ast \bar{D}$, thus the $P_c(4440)$ is expected to decay to the $NJ/\psi$, $\Lambda_c\bar{D}^\ast$, and $\Sigma_c^\ast \bar{D}$ channels via $S$ wave. 

The $P_c$ states with $I(J^P)=1/2(5/2^-)$ can rearrange to $\Sigma_c^\ast \bar{D}^\ast$. Nevertheless, the calculated mass of $P_c(4457)$ is below the threshold of $\Sigma_c^\ast \bar{D}^\ast$. Thus, no $S$-wave decay of $P_c(4457)$ is expected in our diquark model.

The $P_{cs}(4459)$ is assigned to be a hidden charm pentaquark state with isospin spin-parity $IJ^{P}=0\left(1/2^-\right)$ and mass $M=4446$ MeV in model I, while in model II, its quantum numbers are predicted to be $IJ^{P}=0\left(1/2^-\right)$ and $M=4454$ MeV, or $IJ^{P}=0\left(3/2^-\right)$ and $M=4463$ MeV, respectively. If it has $IJ^{P}=0\left(1/2^-\right)$, its wave functions can rearrange to $\Lambda \eta_c$, $\Lambda J/\psi$, $\Lambda_c \bar{D}_s$, $\Xi_c \bar{D}$, $\Lambda_c\bar{D}_s^\ast$, $\Xi'_c \bar{D}$, $\Xi_c \bar{D}^\ast$, $\Xi'_c \bar{D}^\ast$, and $\Xi_c^\ast \bar{D}^\ast$. But as its mass is below the thresholds of $\Xi_c \bar{D}^\ast$, $\Xi'_c \bar{D}^\ast$, and $\Xi_c^\ast \bar{D}^\ast$, it is expected to decay to the other $6$ channels via $S$ wave. If the $P_{cs}(4459)$ has $IJ^{P}=0\left(3/2^-\right)$, its wave functions can rearrange to $\Lambda J/\psi$, $\Lambda_c\bar{D}_s^\ast$, $\Xi_c \bar{D}^\ast$, $\Xi'_c \bar{D}^\ast$, $\Xi_c^\ast \bar{D}$, and $\Xi_c^\ast \bar{D}^\ast$. As its mass is below the thresholds of $\Xi_c \bar{D}^\ast$, $\Xi'_c \bar{D}^\ast$, $\Xi_c^\ast \bar{D}$, and $\Xi_c^\ast \bar{D}^\ast$, it is expected to decay to the $\Lambda J/\psi$ and $\Lambda_c\bar{D}_s^\ast$ channels via $S$ wave.

The rearrangements of the pentaquark wave functions provide us the hints of possible $S$-wave decays. Actually, the $D$-wave decays may also be observed if the mass of a pentaquark state is above the corresponding threshold of the baryon-meson channel whose quantum numbers are allowed. One sees that the $P_c(4312)$ can decay to the $N\eta_c$ and $\Lambda_c {\bar D}$ channels via $D$ wave. The $P_c(4440)$ can decay to the $N\eta_c$, $\Lambda_c {\bar D}$, and $\Sigma_c {\bar D}$ channels via $D$ wave. The $P_c(4457)$ can decay to the $N\eta_c$, $NJ/\psi$, $\Sigma_c {\bar D}$, and $\Sigma_c^\ast {\bar D}$ channels via $D$ wave. Note that the $P_c(4457)$ cannot decay to $\Lambda_c {\bar D}$ and $\Lambda_c {\bar D}^\ast$ via $D$ wave as the two light quarks inside the baryon should have isospin $1$ since they have spin $1$ as shown in Eq.~(\ref{Eq:mass_octet_5_2}). The $P_{cs}(4459)$ with $IJ^{P}=0\left(3/2^-\right)$ can decay to the $\Lambda \eta_c$, $\Lambda_c {\bar D}_s$, $\Xi_c {\bar D}$, and $\Xi'_c {\bar D}$ channels via $D$ wave.

\section{Summary}\label{Sect:Summary}

Although the $P_c$ and $P_{cs}$ states have been widely studied in literature, their quantum numbers are still not clear, as no conclusive answers have been obtained from different approaches or even from the same approach employed in different theoretical works. In the present work, we study the mass spectrum of hidden charm pentaquark states composed of two diquarks and an antiquark by use of an effective Hamiltonian, which includes explicitly the color, spin, and flavor dependent interactions. We construct the color, flavor, and spin wave functions for all possible $(qq)(qc){\bar{c}}$ configurations ($q=u,d,s$), present the mass formulas for all of them, and calculate their masses by use of the model parameters fixed to the known hadron states. 

Our numerical results show that the $P_c(4312)^+$ and $P_c(4440)^+$ states could be explained as hidden charm pentaquark states with isospin and spin-parity $IJ^P=1/2\left(3/2^-\right)$, the $P_c(4457)^+$ state could be explained as a hidden charm pentaquark state with $IJ^P=1/2\left(5/2^-\right)$, and the $P_{cs}(4459)^+$ state could be explained as a hidden charm pentaquark state with $IJ^P=0\left(1/2^-\right)$ or $0\left(3/2^-\right)$. 

The masses of the partners for the $P_c$ and $P_{cs}$ states are also predicted by use of the same set of model parameters. The possible decay channels of the hidden charm pentaquark states are discussed. These states are expected to be searched for in experiments in the near future. They can help to test the diquark model proposed in the present work, and, most hopefully, put further constraints on the model ingredients and parameters.

%

\begin{acknowledgments}
This work is partially supported by the National Natural Science Foundation of China under Grants No.~11475181, No.~11635009, and No.~12075018, the Fundamental Research Funds for the Central Universities, and the Key Research Program of Frontier Sciences of the Chinese Academy of Sciences under Grant No.~Y7292610K1.
\end{acknowledgments}

\end{document}